# Quantum computing of analytical functions by linear optics methods

Mikhail S. Podoshvedov[1,2] and Sergey A. Podoshvedov[1*]

[1]*Laboratory of Quantum Information Processing and Quantum Computing, Institute of Natural and Exact Sciences, South Ural State University (SUSU), Lenin Av. 76, Chelyabinsk, Russia*
[2]*Institute of Physics, Kazan Federal University (KFU), 16a Kremlyovskaya St., Kazan, Russia*
[*]sapodo68@gmail.com

**Abstract:** We propose a model for computing of a certain set of analytical functions based on estimating the output distribution of multiphoton outcomes in an optical scheme with an initial single-mode squeezed vacuum (SMSV) state and photonic states measuring the number of photons in one of the output modes of the beam splitter (BS) by photon number resolving (PNR) detector. The set of considered analytical functions is polynomial expressions including arbitrary derivatives of certain functions which can take on very large values even on small interval in their argument and small values of the parameter indicating the number of the subtracted photons. The large values that the analytic functions can take are offset by a very small term including the factorial of the number of subtracted photons, which guarantees an output normalized distribution of multiphoton measurement outcomes. The quantum computing algorithm makes it possible to find the values of the analytical functions for each number of extracted photons after a sufficiently large number of trials that would allow replacing the measurement repetition rate of multiphoton events by their probabilities. Changing the initial parameters (squeezing amplitude of the SMSV state and BS parameter) makes it possible to implement calculations of the functions over the entire (or, at least, significant) continuous interval of alteration in their argument. The potential of optical quantum computing based on nonclassical states of a certain parity can be expanded both by adding new optical elements such as BSs, and by using other continuous variable (CV) states of definite parity.

## 1. Introduction

Quantum mechanics provides researchers with two fundamental traits superposition and entanglement that allow them to implement quantum computing based on rules other than classical ones [1]. Quantum computing involves parallel computing in which the values of the required function could be calculated at all points of its argument before the superposition is transferred to a fixed base state by measurement [2] with assigning a measured value to a physical quantum computing, thus, giving its estimation, and requires significant repetition of the calculations to obtain estimates of the event probabilities. Quantum computing can be implemented in the gated interpretation [3,4], by creating a cluster state with subsequent measurement in an adjustable basis [5,6], as well as nonuniversal calculations based on sampling model [7,8] when the probability to measure a specific photon pattern on output of $N^2-$interferometer with $N$ input photons depends upon the permanent of some submatrix. The mobility and bosonic nature of photons can be extremely resource-efficient to solve sampling problem that is considered to be classically hard [9]. Another hard-to-calculate function from $\#P$ complexity class Hafnian of an adjacency matrix of a graph can be estimated in Gaussian Boson sampling (GBS) protocol [10] in which SMSV states are utilized as nonclassical resource. Now, only a small number of experiments have been performed that could be considered as computing which go beyond the standard classical simulations [11-14]. Despite the undoubted progress towards quantum computing, quantum supremacy of high-fidelity 53−qubit superconducting quantum processor [11] has been called into question [15], which can undoubtedly only increase interest to the quantum computing and the scientific dispute can only contribute to the further development of the quantum technologies.

Light is undoubtedly the physical medium in which optical quantum computing can be implemented since the speed of propagation of the optical flying qubits is maximum, not depending on the temperature of environment whose influence on their quantum properties is minimal [16]. The advantages could be decisive on the way to creating optical quantum computing if it were not for the fact that photons do not interact directly with each other [17,18], but only through a nonlinear medium [19]. This fact manifests itself in a decrease in the success probability from half to a quarter of trials depending on the protocol used, therefore, the optical flying qubits can hardly be used in the construction of quantum computer blocks in gate architecture. An attempt to enhance efficiency of optical quantum protocols in terms of the success probability leads to a sharp increase in the scalability of quantum optical gates [20], for example, controlled-not gate, and such design of only sole two-qubit element is hardly attractive and feasible in practice. Therefore, the problem of increasing the efficiency of optical qubits interaction [21,22], and, as a consequence, the construction of effective optical quantum gates in linear optics framework is very relevant. Here, we propose a practical scheme of computing analytical functions based on the subtraction of a certain number of photons from the SMSV state. The approach of subtracting small number of photons (say, up to 3 photons) from the initial CV state [23-28], as well as the following technologies following [29-33], is well known. Further development of the method through greater photon subtraction from original nonclassical SMSV state with arbitrary initial conditions [34] makes sense both from a fundamental point of view and from a technological point of view. Thus, quantum engineering of large-amplitude$\geq 5$ high-fidelity$\geq 0.99$ even/odd Schrödinger cat states (SCSs) can acquire an additional impulse [35,36], and the measurement-induced CV states of definite parity realized after the passage of the SMSV through the BS can have a serious metrological potential [37]. The optical scheme for quantum computation of some analytic functions can be fascinatingly simple and based on resources already in use [38]. Measurement-induced mechanism of the CV states of definite parity generation can be sufficient to compute set of the analytical functions after the repetition rate of multiphoton outcomes can be estimated as the probability of an event. The normalized multiphoton distribution allows for one to simultaneously find the values of the set of the analytical functions, the serial number which is determined by the number of subtracted photons. Computing the analytic functions for different values of its argument can be implemented by changing the squeezing amplitude of the original SMSV state and also by varying the transmittance and reflectance of the BS. The simplest analytic functions (arbitrary derivatives of the function $arcsin$) can be calculated in the case of a vacuum state in one of the BS modes. Calculation of more complex analytical functions (polynomial differential functions up to the third order inclusively) can be implemented by adding photonic states (single photon and two-photon state in the case considered) to the input of the BS. We are not talking about quantum superiority of the computing proposed, but taking into account new technology able accurately to distinguish multiphoton states up to 100 photons through their demultiplexing into three measurement channels and further progress in the development of brighter sources of the SMSV states [39], it may be possible to speak about the practical avail of the quantum calculations compared with classical. Indeed, the calculations of the polynomial differential functions by the methods of linear optics are carried out immediately after estimating the distribution of multiphoton outputs, in contrast to the classical composite ones, when, for example, the derivative is first calculated, which is multiplied with some algebraic expression, and so on. In the case of calculating the polynomial differential functions involving very large derivatives (say, 1000 or even more derivatives), the calculation time of the functions with help of linear optics methods can even become less than if they were calculated on a classical computer.

## 2. Computing polynomial differential functions by measurement statistics of multiphoton states

The family the CV states of a certain parity can be realized by subtracting a certain number of photons from the initial SMSV state [34-37]. Measurement-induced mechanism of the CV states generation includes redirection of a part of $n$ photons from original superposition with a complete loss of information about which Fock states the $n$ photons split off to the measurement mode with their measurement mode by PNR detector [34]. The redirection of the $n$ indistinguishable photons into the measurement mode is carried out with help the lossless beam splitter $BS_{12} = \begin{bmatrix} t & -r \\ r & t \end{bmatrix}$, with real transmittance $t$ and reflectance $r$ coefficients satisfying the physical condition $t^2 + r^2 = 1$ and guaranteeing the mixing of modes 1 and 2 by transformation of the creation operators $a_1^+$ and $a_2^+$ as $BS_{12} a_1^+ BS_{12}^+ = ta_1^+ - ra_2^+$ and $BS_{12} a_2^+ BS_{12}^+ = ra_1^+ + ta_2^+$. The BS is considered to be no longer necessarily balanced that means that it can have arbitrary parameters $t$ and $r$ [34]. A single-mode squeezed vacuum state [39-41] occupies first mode of the BS and is described by

$$|SMSV\rangle = \frac{1}{\sqrt{\cosh s}} \sum_{n=0}^{\infty} \frac{y^n}{\sqrt{(2n)!}} \frac{(2n)!}{n!} |2n\rangle, \qquad (1)$$

where $y = \tanh s/2$ [37] and $s > 0$ (squeezing amplitude) is the squeezing parameter of the SMSV that provides the following range of its change $0 \le y \le 0.5$. Value of the squeezing parameter $y = 0$ indicates the absence of the SMSV state at the input of the BS, while the value $y = 0.5$ corresponds to the physically unrealizable case of maximally squeezed light with amplitude $s \to \infty$. In addition to two mutually related SMSV parameters $s$ and $y$, one can also use two other parameters to describe the SMSV state, namely, the squeezing $S$ expressed in $dB$ as $S = -10 \log(\exp(-2s))$ and the mean number of photons $\langle n \rangle_{SMSV} = \sinh^2 s$ in the SMSV state.

Depending on the photonic state in the second input mode, all measurement-induced CV states can also be divided into families depending on the input Fock state used. Indeed, if the second mode is occupied by the number state $|k\rangle$, then the output entangled state can follow from the entangled hybrid state $BS_{12}(|SMSV\rangle_1 |0\rangle_2)$ in Eq. (A1) as

$$BS_{12}(|SMSV\rangle_1 |k\rangle_2) = \frac{1}{\sqrt{k!}} (ra_1^+ + ta_2^+)^k BS_{12}(|SMSV\rangle_1 |0\rangle_2). \qquad (2)$$

Any entangled hybrid states can be derived most simply if the second mode of the BS remains void, that is, in a vacuum state $|0\rangle$. Thus, the output entangled hybrid state in Eq. (A1) can become the basis for the implementation of new entangled hybrid states in Eq. (2), when the photonic state $|k\rangle$ is used in input second mode of the BS. Moreover, the case in Eq. (A1) is of considerable interest from a practical point of view, since it uses a minimum set of resources. In the case of measuring the number of photons which in an indistinguishable manner are diverted to the measuring channel of the beam splitter, heralding generation of either even or odd CV states of certain parity is possible [35-37]. The parity of the measurement-induced CV states is certain, that is, they contain either exclusively even (Eq. (A2)) or odd (Eq. (A3)) Fock states [35-37]. It should be noted that the accurate measurement of number of photons can be produced in a scheme with splitting off multiphoton state (say 100 photons), with their redistribution into three measuring channel realized on base of transition-edge sensors (TESs) [38].

Here, we are not interested in quantum engineering of new CV states of definite parity [36,36] which has serious potential for optical quantum information processing. We use the mechanism of the conditional generation of new CV states of a certain parity in order to estimate the values of set of the analytic functions. The statistics of measurements of the number of photons in the second mode follows entirely from the amplitudes in Eq. (A4) of the output entangled hybrid state in Eq. (A1)

$$P_{2m}^{(0)}(y_1, B) = \frac{\left|c_{2m}^{(0)}(y_1,B)\right|^2}{\cosh s} = \frac{(y_1 B)^{2m}}{\cosh s (2m)!} Z^{(2m)}(y_1) = f_{2m}^{(0)}(y_1, B) Z^{(2m)}(y_1), \quad (3)$$

$$P_{2m+1}^{(0)}(y_1, B) = \frac{\left|c_{2m+1}^{(0)}(y_1,B)\right|^2}{\cosh s} = \frac{(y_1 B)^{2m+1}}{\cosh s (2m+1)!} Z^{(2m+1)}(y_1) = f_{2m+1}^{(0)}(y_1, B) Z^{(2m+1)}(y_1), \quad (4)$$

Here, the probabilities are factorized where one is the $2m, 2m+1$ derivative of the function $Z(y_1) = 1/\sqrt{1-4y_1^2}$ of only one parameter $y_1 = t^2 y$, i.e. $Z^{(2m)}(y_1) = dZ^{2m}/dy_1^{2m}$, $Z^{(2m+1)}(y_1) = dZ^{2m+1}/dy_1^{2m+1}$, while the other additional functions $f_{2m}^{(0)}(y_1, B) = (y_1 B)^{2m}/(\cosh s (2m)!) = \sqrt{1-4y_1^2(1+B)^2} (y_1 B)^{2m}/(2m)!$ and $f_{2m+1}^{(0)}(y_1, B) = \sqrt{1-4y^2} (y_1 B)^{2m+1}/(2m+1)! = \sqrt{1-4y_1^2(1+B)^2} (y_1 B)^{2m+1}/(2m+1)!$ depends on two parameters $y$ and $B$ as $\cosh s = 1/\sqrt{1-4y^2} = 1/\sqrt{1-4y_1^2(1+B)^2}$, where the beam splitter parameter (BSP) $B = (1-t^2)/t^2$ is introduced. The parameter $y_1$ can vary in the range of $0 < y_1 \le 0.5/(1+B)$ which ensures the fulfillment of the condition $y_1 \le y$. Here, the additional functions $f_{2m}^{(0)}(y_1, B), f_{2m+1}^{(0)}(y_1, B)$ respectively, and the probabilities themselves in Eqs. (3,4) depend on $y_1$ and $B$ which determine the squeezing parameter $y = (1+B)y_1$ and, as consequence, squeezing amplitude $s$ of the original SMSV in Eq. (1). On the contrary, knowing the squeezing amplitude $s$ of the initial SMSV state and the transmission $T = t^2$ of BS used, one can proceed to the parameters $B$ and $y_1 = y/(1+B)$. Thus, knowing input values of the parameters $s$ and $t$, the values of the additional functions $f_{2m}^{(0)}(y_1, B)$ and $f_{2m+1}^{(0)}(y_1, B)$ are calculated in advance that guarantees evaluation of values of the following set of the analytical functions

$$Z^{(2m)}(y_1) = \frac{P_{2m}^{(0)}(y_1,B)}{f_{2m}^{(0)}(y_1,B)}, \quad (5)$$

$$Z^{(2m+1)}(y_1) = \frac{P_{2m+1}^{(0)}(y_1,B)}{f_{2m+1}^{(0)}(y_1,B)}. \quad (6)$$

at corresponding point $y_1$, provided that a sufficient number of measurement outcomes is collected to evaluate the measurement probability distribution. By changing the initial parameters of the experiment, that is, $s$ and $t$, thereby changing the value of the parameter $y_1$, the $2m, 2m+1$ derivatives of the function $Z(y_1)$ are already computed at another point $y_1$. In general, continuous domain of $y_1$ can be provided by changing only the values of $s$ leaving the value of the BSP $B$ to be constant, which guarantees the continuity of derivatives of $Z(y_1)$ on the chosen domain of $y_1$ or $s$. Note that $Z(y_1) = d(arcsin(2y_1))/(2dy_1)$, therefore, the $2m, 2m+1$ derivatives of the function $Z(y_1)$ are the $2m+1, 2m+2$ derivatives of the $arcsin$ function.

The approach is not limited only to the computing the derivatives of the $arcsin$ function and can be extended to compute polynomial differential operators of a certain order with respect to the differential operator $y_1 d/dy_1$ acting on some analytical function involving the function $Z(y_1)$ and its derivatives. The formation of the polynomial differential functions is realized by adding additional input photonic states to the second mode of the beam splitter in Fig. 1 that are mixed with the original SMSV state. Let us consider the influence of an additional input single photon on shaping of a set of calculated analytical functions. Output photonic statistics directly follows from the expressions (B7-B9) and gives the following distribution of the measurement outcomes probabilities

$$P_0^{(1)}(y_1, B) = \frac{\left|C_0^{(1)}(y_1,B)\right|^2}{\cosh s} = \sqrt{1 - 4y_1^2(1+B)^2}\frac{B}{1+B}Z^3(y_1) = f_0^{(1)}(y_1,B)Z^3(y_1), \quad (7)$$

$$P_{2m}^{(1)}(y_1, B) = \frac{\left|C_{2m}^{(1)}(y_1,B)\right|^2}{\cosh s} = \frac{\sqrt{1-4y_1^2(1+B)^2}}{1+B}\frac{(y_1B)^{2m-1}}{(2m)!}(2m)^2 G_{2m}^{(1)}(y_1,B) =$$
$$f_{2m}^{(1)}(y_1,B)G_{2m}^{(1)}(y_1,B), \quad (8)$$

$$P_{2m+1}^{(1)}(y_1, B) = \frac{\left|C_{2m+1}^{(1)}(y_1,B)\right|^2}{\cosh s} = \frac{\sqrt{1-4y_1^2(1+B)^2}}{1+B}\frac{(y_1B)^{2m}}{(2m+1)!}(2m+1)^2 G_{2m+1}^{(1)}(y_1,B) =$$
$$f_{2m+1}^{(1)}(y_1,B)G_{2m+1}^{(1)}(y_1,B), \quad (9)$$

where the additional analytical functions $f_0^{(1)}(y_1, B)$, $f_{2m}^{(1)}(y_1, B)$ and $f_{2m+1}^{(1)}(y_1, B)$ follow from Eqs. (B10-B12). Thus, if the initial values of the parameters $s$ and $t$ are known, then the values of $y_1, B$ are determined that enable to calculate the values of the additional functions in advance and by evaluating the measurement outcomes distribution, after a sufficient number of trials, compute the values of the polynomial differential functions of first order as

$$Z^3(y_1) = \frac{P_0^{(1)}(y_1,B)}{f_0^{(1)}(y_1,B)}, \quad (10)$$

$$G_{2m}^{(1)}(y_1, B) = \frac{P_{2m}^{(1)}(y_1,B)}{f_{2m}^{(1)}(y_1,B)}, \quad (11)$$

$$G_{2m+1}^{(1)}(y_1, B) = \frac{P_{2m+1}^{(1)}(y_1,B)}{f_{2m+1}^{(1)}(y_1,B)}. \quad (12)$$

Change of values of the input parameters $s, t$ expands domain of $y_1$ and, thereby, allows for one continuously to compute the values of set of polynomial differential functions of first order on the continuous two-dimensional interval of $y_1$ and $B$. If the calculation task is to compute the function without changing its coefficients, then one needs to change only the value of the parameter s, which is not affected by the change in the parameter $B$.

The polynomial differential functions of a higher order (say, of the third order) are calculated in the case of already adding a two-photon state to the SMSV state and mixing them on BS in the scheme in Fig. 1. Using the mathematical approach presented in Appendix B, we compute the following probability distribution of the measurement outcomes

$$P_0^{(2)}(y_1, B) = \frac{\left|C_0^{(2)}(y_1,B)\right|^2}{2\cosh s} = \frac{\sqrt{1-4y_1^2(1+B)^2}}{2}\left(\frac{B}{2y_1(1+B)}\right)^2\left(y_1\frac{d}{dy_1}\right)\left(y_1 Z^{(1)}(y_1)\right) =$$
$$f_0^{(2)}(y_1,B)\left(y_1\frac{d}{dy_1}\right)\left(y_1 Z^{(1)}(y_1)\right), \quad (13)$$

$$P_{2m}^{(1)}(y_1, B) = \frac{\left|C_1^{(2)}(y_1,B)\right|^2}{2\cosh s} = \sqrt{1-4y_1^2(1+B)^2}\frac{2B}{y_1(1+B)^2}G_1^{(2)}(y_1,B) =$$
$$f_1^{(2)}(y_1,B)G_{2m}^{(1)}(y_1,B), \quad (14)$$

$$P_{2m}^{(2)}(y_1, B) = \frac{\left|C_{2m}^{(2)}(y_1,B)\right|^2}{2\cosh s} = \frac{\sqrt{1-4y_1^2(1+B)^2}}{2(1+B)^2}\frac{(y_1B)^{2m-2}}{(2m)!}(2m-1)^2(2m)^2 G_{2m}^{(2)}(y_1,B) =$$
$$f_{2m}^{(2)}(y_1,B)G_{2m}^{(2)}(y_1,B), \quad (15)$$

$$P_{2m+1}^{(2)}(y_1, B) = \frac{\left|C_{2m+1}^{(2)}(y_1,B)\right|^2}{2\cosh s} = \frac{\sqrt{1-4y_1^2(1+B)^2}}{2(1+B)^2}\frac{(y_1B)^{2m-1}}{(2m+1)!}(2m)^2(2m+1)^2 G_{2m+1}^{(2)}(y_1,B) =$$
$$f_{2m+1}^{(2)}(y_1,B)G_{2m}^{(2)}(y_1,B), \quad (16)$$

where the additional functions $f_0^{(2)}(y_1, B), f_1^{(2)}(y_1, B)$  $f_{2m}^{(2)}(y_1, B)$ and $f_{2m+1}^{(2)}(y_1, B)$ are presented in Eqs. (B26-B29). Again the initial values of $s$ and $t$ determine the value of the parameters $y_1$ and $B$. Knowing the probability distribution of the measurement outcomes, we can calculate polynomial differential functions of third-order at some specific point $y_1$ as

$$\left(y_1 \frac{d}{dy_1}\right)\left(y_1 Z^{(1)}(y_1)\right) = \frac{P_0^{(2)}(y_1,B)}{f_0^{(2)}(y_1,B)}, \tag{17}$$

$$G_1^{(2)}(y_1,B) = \frac{P_1^{(2)}(y_1,B)}{f_1^{(2)}(y_1,B)}, \tag{18}$$

$$G_{2m}^{(2)}(y_1,B) = \frac{P_{2m}^{(2)}(y_1,B)}{f_{2m}^{(2)}(y_1,B)}, \tag{19}$$

$$G_{2m+1}^{(2)}(y_1,B) = \frac{P_{2m+1}^{(2)}(y_1,B)}{f_{2m+1}^{(2)}(y_1,B)}. \tag{20}$$

The single point calculation can be extended to a continuous two-dimensional domain of $y_1$ and $B$ by corresponding variation of the input parameters $s$ and $t$. To trace the dynamics of the same function only on the interval $y_1$, it is worth varying only the parameter $s$, the change of which does not affect its coefficients.

In order to present the possibilities of quantum computing based on the redirection of the initial photons into the measuring mode so that the information from which Fock state they could be split off is completely lost, we show in figure 2 the dependences of the derivatives of the function $Z(y_1)$ on the parameter $y_1$ for different values of the parameter $m$. The range of the argument $y_1$ in figure 2 is chosen to be quite small, i.e. for $0 \leq y_1 \leq 0.2$, which does not prevent the $n-$derivatives from taking very large values even for moderate values of the parameter $m \leq 47$. Even a slight increase in both the variable $y_1 > 0.2$ and the parameter $m > 47$ leads to an even sharper increase in the values of the derivatives, so, this increase may be so significant that a classic personal computer may already spend a significant amount of time calculating them. Approximately the same sharp increase in values is observed for the polynomial differential functions of the first in Eqs. (B5,B6) and third order in Eqs. (B18-B21), even in an even smaller range of changes in the variable $y_1 \leq 0.05$ at comparative values of the parameter $m$. This sharp increase in the values of the polynomial differential functions is compensated by a term like $(y_1 B)^n/n!$ which can take on very small values. The compensating effect of the term is sufficient for their product to take acceptable values, which, in combination with the remaining terms, guarantees a normalized multiphoton distribution. Carrying out a sufficiently large number of measurements allows us to estimate the frequency of occurrence of certain measurement events, which, in the case of a sufficiently large number of trials, allows us to replace them on their probabilities. We also note that each measurement outcome can not only contribute to the frequency of occurrence of the corresponding measurement event, but also generate new conditional CV states of certain parity in Eqs. (A2,A3,B2-B4,B14-B17), which could also be used in optical quantum information processing [35-37].

## 3. Conclusion

For the first time to our knowledge, we have proposed a model for calculating analytic functions in a manner different from the previous based on gate architecture, cluster states, and also on the bosonic sampling model. The model with subtraction of some number of photons from original SMSV state can be used to calculate the values of polynomial differential functions at certain points of $y_1$ and $B$. It allows for one to calculate the values of the functions as a whole without resorting to the calculation of its constituent parts with their subsequent manipulations (addition, subtraction, etc.) as it is implemented in classical calculation. In addition to the dependence on $s$ and $t$, the functions to a large extent depend on the number $2m, 2m+1$ of extracted photons. So, the number $m$ can determine the derivative of the function $Z(y_1)$ to which the polynomial differential operator is applied. Experimental obtaining the measurement outcomes distribution depending on the number of extracted photons allows for one to calculate the set of the corresponding polynomial differential functions for each an integer parameter $m$ that ideally takes values in the range from 0 to ∞. Expansion of the computing for a continuous two-dimensional interval of values $y_1$ and $B$ can be implemented by varying the initial parameters of the optical setup $s$ and $t$ that changes $y_1$ and $B$. In general, the calculation of the polynomial differential functions at various points from its domain of definition of $y_1$ occurs at constant values of their coefficients, which requires a constant value of $t = const$ and, as a result, $B = const$ that imposes a serious condition for creating a configurable source of the SMSV states. The probability of extracting a large number of photons from the SMSV state with a small mean number of photons $\langle n \rangle_{SMSV} < 1$ is very small. Therefore, use of the original SMSV state with $s \ll 1$ may limit usefulness of the treatment to implement useful computing in practice as considered polynomial differential functions only with small number $m$ can be calculated due to the practical impossibility of correct estimating a very small probability value for large values of $m$. The same polynomial differential functions with small values of $m$ can be evaluated on a classical computer in a reasonable amount of time. The development of a technology for generating a bright SMSV state [39] (even desirable with $\langle n \rangle_{SMSV} > 100$) makes it possible to increase the probability of detecting multiphoton states (say, ≥ 100 photons), which, in combination with the development of accurate photon number resolving technique [38] based on their demultiplication towards TESs channels, can become appropriate for calculating polynomial differential functions after practical estimation of the distribution of multiphoton states. In the case of developing a technology for detecting larger number of photons with a set of TESs (for example, capable of detecting 1000 photons or more), it may even be possible to talk about quantum superiority realized by linear optics methods. In the case of extracting a small number of photons (say, 10 photons), quantum superiority cannot be mentioned. Note that we have considered the calculation of the polynomial differential functions in Eqs. (5,6,B5,B6,B18-B21) provided that additional functions in Eqs. (B10-B12,B26-B29) have already been calculated. The polynomial differential functions can take on very large values while the additional functions can take on very small values. Multiplying a very large number by a very small number gives a probability of a measurement outcome whose value is less than $< 1$. Instead of pre-calculating the additional functions that take very small values, one can consider their product as the goals of the calculations. For example, in the case of input two-photon state with even measurement outcome, one can calculate the value of the following function $((y_1 B)^{2m-2}/(2m)!)G^{(2)}_{2m}(y_1, B)$ which already takes on significantly smaller values compared with initial $G^{(2)}_{2m}(y_1, B)$ by precomputing only simple expression $\left(1 - 4y_1^2(1+B)^2(2m(2m-1))^2\right)/2(1+B)^2$. Note that the quantum computations are not applicable to all analytic functions, but only to certain polynomial differential functions definable by generated CV state of a certain parity which is not in favor of the universality of the calculations. Despite the fact that the model of quantum calculations based on generation of new CV states of definite parity by subtraction of arbitrary number of photons from SMSV state can only compute only a certain set of functions on some continuous domain, the importance of the extended model can only grow. An extended version of the quantum computing could include finding the values of already other analytic functions by using new CV states of a certain parity and Fock states at the input. So even use of additional photonic states at the input to the BS makes it possible to calculate the first order polynomial differential functions in the case of including input single photon instead of just derivatives of the functions $Z(y_1)$ and the ones of third order in the case of using an additional input two-photon state. Extended approach may be related to the implementation of a set of beam splitters, which are fed with various nonclassical states as CV and photonic states which can lead to the calculation of a different set of analytic functions provided that output multiphoton statistics can be correctly estimated.

## Appendix A. Entangled hybrid state at the output of the BS and measurement-induced generation of CV states of definite parity

The SMSV state in Eq. (1) with input amplitude $y$ is converted into entangled hybrid state

$$BS_{12}(|SMSV\rangle_1|0\rangle_2) = \frac{1}{\sqrt{\cosh s}}\sum_{l=0}^{\infty} C_l^{(0)}(y_1, B)\, |\Psi_l^{(0)}(y_1)\rangle_1 |l\rangle_2, \qquad (A1)$$

where even $l = 2m$ CV states of definite parity are the following

$$|\Psi_{2m}^{(0)}(y_1)\rangle = \frac{1}{\sqrt{Z^{(2m)}(y_1)}}\sum_{n=0}^{\infty} \frac{y_1^n}{\sqrt{(2n)!}} \frac{(2(n+m))!}{(n+m)!} |2n\rangle, \qquad (A2)$$

while the odd $l = 2m + 1$ CV states of certain parity are turned out to be

$$|\Psi_{2m+1}^{(0)}(y_1)\rangle = \sqrt{\frac{y_1}{Z^{(2m+1)}(y_1)}}\sum_{n=0}^{\infty} \frac{y_1^n}{\sqrt{(2n+1)!}} \frac{(2(n+m+1))!}{(n+m+1)!} |2n+1\rangle, \qquad (A3)$$

and the input SMSV parameter $y$ is reduced by a multiplier $t^2$ with respect to initial value $y_1 = t^2 y \le y$. Definition of the parameters used is presented in the main part of the manuscript. As for the indices used here and elsewhere, the superscript indicates the number of photons entering the BS while the subscript shows the number of extracted photons from the original SMSV state. The amplitudes $C_l^{(0)}(y, y_1)$ of the entangled hybrid state (A1) become

$$C_l^{(0)}(y_1, B) = (-1)^l \frac{(By_1)^{\frac{l}{2}}}{\sqrt{l!}} \begin{cases} \sqrt{Z^{(2m)}(y_1)}, & \text{if } l = 2m \\ \sqrt{Z^{(2m+1)}(y_1)}, & \text{if } l = 2m+1 \end{cases}, \qquad (A4)$$

where the beam splitter parameter $B$ is introduced above.

## Appendix B. Influence of input photonic states on output entangled hybrid states and families of the measurement-induced CV states of definite parity

A new entangled hybrid state follows if a single photon is added at the input to the second mode of the BS that leads to new hybrid entangled state

$$BS_{12}(|SMSV\rangle_1|1\rangle_2)\frac{1}{\sqrt{\cosh s}}\begin{pmatrix} C_0^{(1)}(y_1,B)|\Psi_0^{(1)}(y_1)\rangle_1 |0\rangle_2 - \\ -\sum_{m=1}^{\infty} C_{2m}^{(0)}(y_1,B)\,|\Psi_{2m}^{(1)}(y_1,B)\rangle_1 |2m\rangle_2 + \\ +\sum_{m=0}^{\infty} C_{2m+1}^{(0)}(y_1,B)\,|\Psi_{2m+1}^{(1)}(y_1,B)\rangle_1 |2m+1\rangle_2 \end{pmatrix}, \qquad (B1)$$

where the odd CV states with subscript $2m$ are given by

$$|\Psi_0^{(1)}(y_1)\rangle = \frac{1}{\sqrt{Z^3(y_1)}}\sum_{n=0}^{\infty} \frac{y_1^n}{\sqrt{(2n+1)!}} \frac{(2n+1)!}{n!} |2n+1\rangle, \qquad (B2)$$

for $m = 0$

$$|\Psi_{2m}^{(1)}(y_1,B)\rangle = \sqrt{\frac{y_1}{G_{2m}^{(1)}(y_1,B)}}\sum_{n=0}^{\infty} \frac{y_1^n}{\sqrt{(2n+1)!}} \frac{(2(n+m))!}{(n+m)!}\left(1 - B\frac{2n+1}{2m}\right)|2n+1\rangle, \qquad (B3)$$

for $m \ne 0$, while the even CV states with subscript $2m + 1$ are determined by

$$|\Psi_{2m+1}^{(1)}(y_1,B)\rangle = \frac{1}{\sqrt{G_{2m+1}^{(1)}(y_1,B)}}\sum_{n=0}^{\infty} \frac{y_1^n}{\sqrt{(2n)!}} \frac{(2(n+m))!}{(n+m)!}\left(1 - B\frac{2n}{2m+1}\right)|2n\rangle, \qquad (B4)$$

whose normalization factors are the following

$$G_{2m}^{(1)}(y_1, B) = Z^{(2m-1)}(y_1) - \frac{B}{m}\left(y_1 Z^{(2m)}(y_1)\right) + \left(\frac{B}{2m}\right)^2 y_1 \frac{d}{dy_1}\left(y_1 Z^{(2m)}(y_1)\right) =$$
$$Z^{(2m-1)}(y_1) + L_{2m}^{(1)}(y_1, B)\left(y_1 Z^{(2m)}(y_1)\right), \tag{B5}$$

$$G_{2m+1}^{(1)}(y_1, B) = Z^{(2m)}(y_1) - \frac{2B}{2m+1}\left(y_1 Z^{(2m+1)}(y_1)\right) + \left(\frac{B}{2m+1}\right)^2 y_1 \frac{d}{dy_1}\left(y_1 Z^{(2m+1)}(y_1)\right) =$$
$$Z^{(2m)}(y_1) + L_{2m+1}^{(1)}(y_1, B)\left(y_1 Z^{(2m+1)}(y_1)\right), \tag{B6}$$

here the linear differential operators $L_{2m}^{(1)}(y_1, B)$ and $L_{2m+1}^{(1)}(y_1, B)$ acting on the analytical functions either $\left(y_1 Z^{(2m)}(y_1)\right)$ or $\left(y_1 Z^{(2m+1)}(y_1)\right)$ with corresponding amplitudes are introduced. Note that here and throughout the work we use the following notation for the differential operator acting on the analytical function $\left(y_1 Z^{(l)}(y_1)\right)$, i.e. $(y_1 d/dy_1)^{(n)}\left(y_1 Z^{(l)}(y_1)\right) = y_1 d/dy_1 \left((y_1 d/dy_1) \ldots (y_1 d/dy_1)\left(y_1 Z^{(l)}(y_1)\right)\right)$, where the derivative procedure $y_1 d/dy_1$ is repeated $n$ times. Amplitudes of the entangled hybrid state in Eq. (B1) that determine the measurement outcome probability distribution and probabilities to generate measurement-induced CV states of certain parity in Eqs. (B2-B4) are given by

$$C_0^{(1)}(y_1, B) = \sqrt{\frac{B}{1+B}} Z^3(y_1), \tag{B7}$$

$$C_{2m}^{(1)}(y_1, B) = \frac{1}{\sqrt{1+B}} \frac{(y_1 B)^{m-\frac{1}{2}}}{\sqrt{(2m)!}} (2m) \sqrt{G_{2m}^{(1)}(y_1, B)}, \tag{B8}$$

$$C_{2m+1}^{(1)}(y_1, B) = \frac{1}{\sqrt{1+B}} \frac{(y_1 B)^m}{\sqrt{(2m+1)!}} (2m+1) \sqrt{G_{2m+1}^{(1)}(y_1, B)}. \tag{B9}$$

Knowledge of the amplitudes of the hybrid entangled state allows us to derive the functions $f_0^{(1)}(y_1, B)$, $f_{2m}^{(1)}(y_1, B)$ and $f_{2m+1}^{(1)}(y_1, B)$ as

$$f_0^{(1)}(y_1, B) = \sqrt{1 - 4y_1^2(1+B)^2} \frac{B}{1+B}, \tag{B10}$$

$$f_{2m}^{(1)}(y_1, B) = \frac{\sqrt{1-4y_1^2(1+B)^2}}{1+B} \frac{(y_1 B)^{2m-1}}{(2m)!} (2m)^2, \tag{B11}$$

$$f_{2m+1}^{(1)}(y_1, B) = \frac{\sqrt{1-4y_1^2(1+B)^2}}{1+B} \frac{(y_1 B)^{2m}}{(2m+1)!} (2m+1)^2. \tag{B12}$$

Let us extend the previous analysis to include the interaction of SMSV state with a two-photon state which leads to realization of the following entangled hybrid state

$$BS_{12}(|SMSV\rangle_1 |2\rangle_2) = \frac{1}{\sqrt{2 \cosh s}}$$
$$\begin{pmatrix} C_0^{(2)}(y_1, B)|\Psi_0^{(2)}(y_1)\rangle_1 |0\rangle_2 + C_1^{(2)}(y_1, B)|\Psi_1^{(2)}(y_1, B)\rangle_1 |1\rangle_2 + \\ \sum_{m=1}^{\infty} C_{2m}^{(2)}(y_1, B) |\Psi_{2m}^{(2)}(y_1, B)\rangle_1 |2m\rangle_2 - \sum_{m=1}^{\infty} C_{2m+1}^{(2)}(y_1, B) |\Psi_{2m+1}^{(2)}(y_1, B)\rangle_1 |2m+1\rangle_2 \end{pmatrix}, \tag{B13}$$

where

$$|\Psi_0^{(2)}(y_1)\rangle = \frac{2}{\sqrt{\left(y_1 \frac{d}{dy_1}\right)\left(y_1 Z^{(1)}(y_1)\right)}} \sum_{n=0}^{\infty} \frac{y_1^n}{\sqrt{(2n)!}} \frac{(2n)!}{n!} n |2n\rangle, \tag{B14}$$

$$|\Psi_1^{(2)}(y_1, B)\rangle = \sqrt{\frac{y_1}{G_1^{(2)}(y_1, B)}} \sum_{n=0}^{\infty} \frac{y_1^n}{\sqrt{(2n+1)!}} \frac{(2n)!}{n!} (2n+1)(1-Bn)|2n+1\rangle, \tag{B15}$$

$$|\Psi_{2m}^{(2)}(y_1, B)\rangle = \frac{1}{\sqrt{G_{2m}^{(2)}(y_1, B)}} \sum_{n=0}^{\infty} \frac{y_1^n}{\sqrt{(2n)!}} \frac{(2(n+m-1))!}{(n+m-1)!}\left(1 - 2B\frac{2n}{2m-1} + B^2 \frac{n(2n-1)}{m(2m-1)}\right)|2n\rangle, \tag{B16}$$

$$|\Psi_{2m+1}^{(2)}(y_1, B)\rangle = \sqrt{\frac{y_1}{G_{2m+1}^{(2)}(y_1, B)}} \sum_{n=0}^{\infty} \frac{y_1^n}{\sqrt{(2n+1)!}} \frac{(2(n+m))!}{(n+m)!}$$
$$\left(1 - 2B\frac{2n+1}{2m} + B^2 \frac{n(2n+1)}{m(2m+1)}\right) |2n+1\rangle, \tag{B17}$$

with the following normalization factors for the CV states with $m = 0$ in Eqs. (B14,B15)

$$G_0^{(2)}(y_1) = \frac{1}{4}\left(y_1 \frac{d}{dy_1}\right)\left(y_1 Z^{(1)}(y_1)\right) = L_0^{(2)}(y_1)\left(y_1 Z^{(1)}(y_1)\right), \tag{B18}$$

$$G_1^{(2)}(y_1, B) = \left(1 + B + \frac{B^2}{4}\right)\left(y_1 \frac{d}{dy_1}\right)(y_1 Z(y_1)) - B\left(1 + \frac{B}{2}\right)\left(y_1 \frac{d}{dy_1}\right)^2 (y_1 Z(y_1)) +$$
$$\left(\frac{B}{2}\right)^2 \left(y_1 \frac{d}{dy_1}\right)^3 (y_1 Z(y_1)) = L_1^{(2)}(y_1, B)\left(y_1 Z^{(1)}(y_1)\right), \tag{B19}$$

$$G_{2m}^{(2)}(y_1, B) = Z^{(2m-2)}(y_1) - \frac{4B}{2m-1}\left(1 + \frac{B}{4m}\right)\left(y_1 Z^{(2m-1)}(y_1)\right) +$$
$$\frac{4B^2}{(2m-1)^2}\left(1 + \frac{B^2}{16m^2} + \frac{2m-1}{4m} + \frac{B}{2m}\right)\left(y_1 \frac{d}{dy_1}\right)\left(y_1 Z^{(2m-1)}(y_1)\right) -$$
$$\frac{2B^3}{m(2m-1)^2}\left(1 + \frac{B}{4m}\right)\left(y_1 \frac{d}{dy_1}\right)^2 \left(y_1 Z^{(2m-1)}(y_1)\right) + \frac{B^4}{4m^2(2m-1)^2}\left(y_1 \frac{d}{dy_1}\right)^3 \left(y_1 Z^{(2m-1)}(y_1)\right) =,$$
$$Z^{(2m-2)}(y_1) + L_{2m}^{(2)}(y_1, B)\left(y_1 Z^{(2m-1)}(y_1)\right), \tag{B20}$$

$$G_{2m+1}^{(2)}(y_1, B) = Z^{(2m-1)}(y_1) - \frac{2B}{m}\left(1 + \frac{B}{2(2m+1)}\right)\left(y_1 Z^{(2m)}(y_1)\right) +$$
$$\frac{B^2}{m^2}\left(1 + \frac{B^2}{4(2m+1)^2} + \frac{m}{2m+1} + \frac{B}{2m+1}\right)\left(y_1 \frac{d}{dy_1}\right)\left(y_1 Z^{(2m)}(y_1)\right) -$$
$$\frac{B^3}{m^2(2m+1)}\left(1 + \frac{B}{2(2m+1)}\right)\left(y_1 \frac{d}{dy_1}\right)^2 \left(y_1 Z^{(2m)}(y_1)\right) + \frac{B^4}{(2m)^2(2m+1)^2}\left(y_1 \frac{d}{dy_1}\right)^3 \left(y_1 Z^{(2m)}(y_1)\right) =$$
$$Z^{(2m-1)}(y_1) + L_{2m+1}^{(2)}(y_1, B)\left(y_1 Z^{(2m)}(y_1)\right). \tag{B21}$$

Here, the linear differential operators $L_0^{(2)}(y_1)$, $L_1^{(2)}(y_1, B)$, $L_{2m}^{(2)}(y_1, B)$ and $L_{2m+1}^{(2)}(y_1, B)$ with coefficients following from polynomial differential functions in Eqs. (B18-B21) are also introduced into consideration. The operators act on the analytical functions $y_1 Z^{(1)}(y_1)$, $y_1 Z^{(2m-1)}(y_1)$ and $y_1 Z^{(2m)}(y_1)$, respectively. Amplitudes of the hybrid entangled state (B13) are the following

$$C_0^{(2)}(y_1, B) = \frac{B}{2y_1(1+B)}\sqrt{\left(y_1 \frac{d}{dy_1}\right)\left(y_1 Z^{(1)}(y_1)\right)}, \tag{B22}$$

$$C_1^{(2)}(y_1, B) = \frac{2}{1+B}\sqrt{\frac{B}{y_1}}\sqrt{G_1^{(2)}(y_1, B)}, \tag{B23}$$

$$C_{2m}^{(2)}(y_1, B) = \frac{1}{1+B}\frac{(y_1 B)^{m-1}}{\sqrt{(2m)!}}(2m-1)2m\sqrt{G_{2m}^{(2)}(y_1, B)}, \tag{B24}$$

$$C_{2m+1}^{(2)}(y_1, B) = \frac{1}{1+B}\frac{(y_1 B)^{m-1/2}}{\sqrt{(2m+1)!}}2m(2m+1)\sqrt{G_{2m+1}^{(2)}(y_1, B)}. \tag{B25}$$

Expressions (B22-B25) are used to calculate additional analytic functions as

$$f_0^{(2)}(y_1, B) = \frac{\sqrt{1-4y_1^2(1+B)^2}}{2}\left(\frac{B}{2y_1(1+B)}\right)^2, \tag{B26}$$

$$f_1^{(2)}(y_1, B) = \sqrt{1-4y_1^2(1+B)^2}\frac{2B}{(1+B)^2}, \tag{B27}$$

$$f_{2m}^{(2)}(y_1, B) = \frac{\sqrt{1-4y_1^2(1+B)^2}}{2(1+B)^2}\frac{(y_1 B)^{2m-2}}{(2m)!}(2m-1)^2(2m)^2, \tag{B28}$$

$$f_{2m+1}^{(2)}(y_1, B) = \frac{\sqrt{1-4y_1^2(1+B)^2}}{2(1+B)^2}\frac{(y_1 B)^{2m-1}}{(2m+1)!}(2m)^2(2m+1)^2. \tag{B29}$$

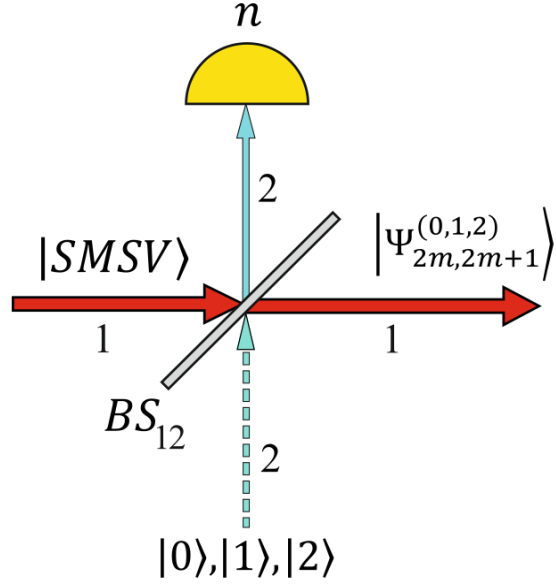

**Figure 1** An optical scheme is used to implement quantum calculations of sets of polynomial differential functions involving derivatives of the function $Z(y_1)$. By making a large number of measurements so that the event repetition rates could be interpreted as a distribution of multiphoton outcomes, one can obtain the value of the corresponding polynomial differential functions at one corresponding point $y_1$. Each set of calculated functions is determined by the input additional photonic state directed to the second auxiliary mode to mix with original SMSV state. Domain of the function argument $y_1$ is determined by the squeezing amplitude $s$ of the initial SMSV state and also by the BSP $B$ which inevitably reduces the input domain of definition of the functions.

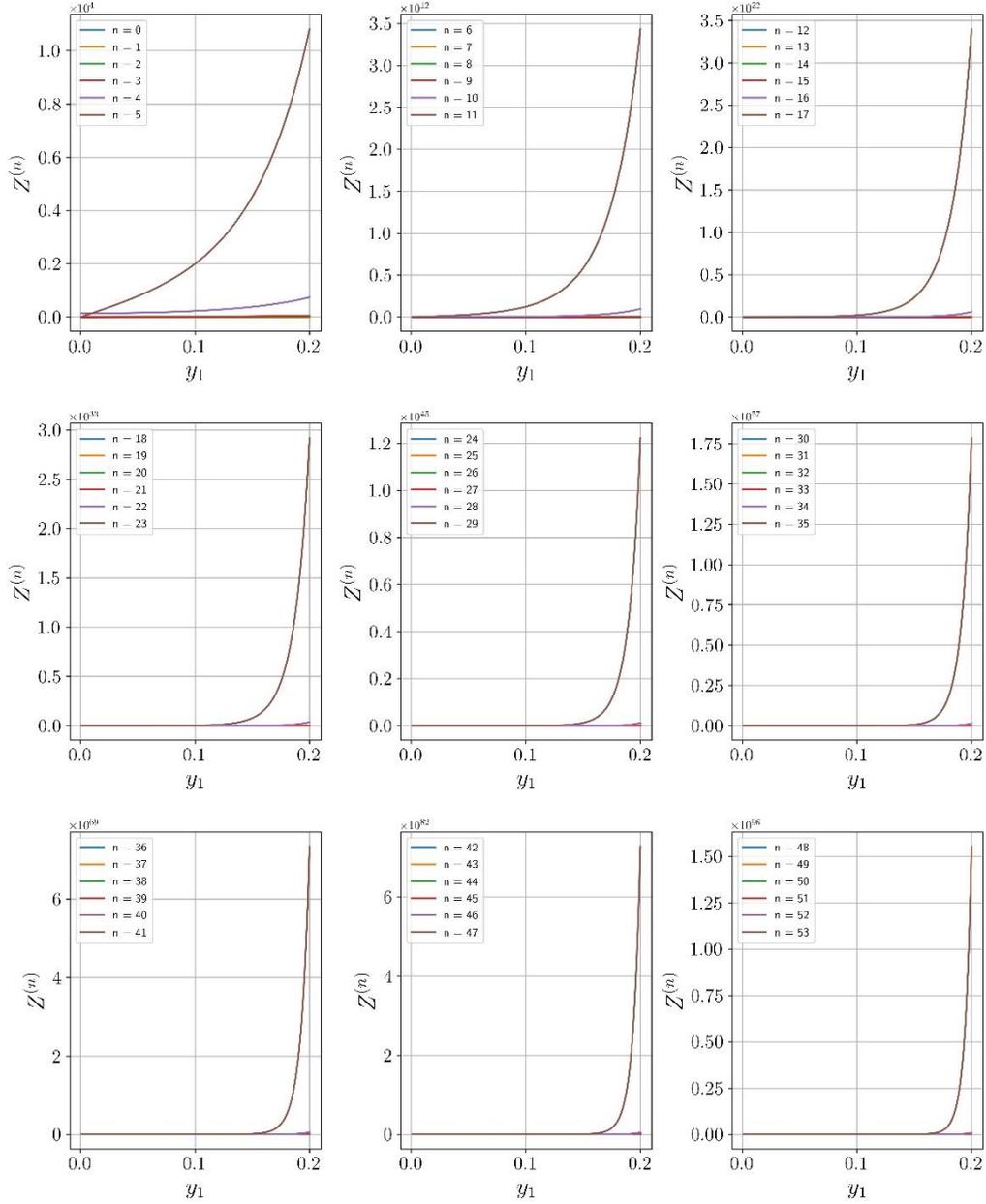

**Figure 2** Dependence of the derivatives of the function $Z^{(n)}(y_1)$ on the variable $y_1$ for different values of the parameter $m$. The derivatives grow very quickly, both depending on the variable $y_1$ and on the order of the derivative that is determined by number of the subtracted photons, and can take very large values even for quite small values of $m$, while the parameter $y_1$ can take on those values that could be implemented in practice i.e. for $y_1 \in [0,0.2]$.

14